# Phonon tunnels from a sonic black hole

H. Z. Fang[*] and K. H. Zhou[†]

*Department of Physics, China University of Petroleum, Changjiang Road 66, 266555, Qingdao, China*

We investigate the phonon radiation from a spherically symmetrical, stationary, viscid-free sonic black hole by using a semi-classical method. The backreaction of the radiated phonon is taken into account. We obtain the phonon emission temperature, and it is consistent with the Hawking's formula.

## I. INTRODUCTION

Hawking's prediction that black hole radiates particles from its horizon with a Planckian spectrum [1] is quite attractive and far-reaching-influenced in theoretical physics field. There are numerous derivations of the Hawking effect which emphasis different features of the process, and sometimes make markedly different physical assumptions. The Hartle-Hawking approach used analytic continuation of the propagator across the event horizon of an eternal black hole [2]. Damour and Ruffini emphasized that the outgoing modes suffer a barrier when crossing the horizon to yield the same Hawking temperature [3]. Robinson and Wilczek claimed that there exists relationship between Hawking radiation and Gravitational Anomalies [4] and some physicists believe that Hawking effect can be interpreted in string theory [5]. In recently years, one of the most interesting works on Hawking radiation derivation done by Parikh and Wilczek [6] comes to public's attention. Parikh and Wilczek believe that black hole radiation do can be regarded as particle tunneling, but they gave a different description of quantum tunneling through the horizon in the process of black hole evaporation. The radiation spectrum they obtained deviates from exact thermal and they proved that the quantum information conserves in the process of black hole evaporation.

As a very important way in physics and mathematics, analogies provide new ways of looking at problems that permit cross-fertilization of ideas among different branches of science. Since the study directly on Hawking radiation from naturally formed black holes, as is known, there are still too many difficulties today, so the analogy of gravity becomes an alternative way. The most beautiful example of this was pointed out by Unruh [7] in 1981. He gave an analogy of light waves in curved spacetime with sound waves viewed as a disturbance of fluid background field in flat Minkowski spacetime, then he deduced an "effective" metric to describe the sound propagation in fluid flow and obtained the phonon Hawking emission from sonic horizon formed in transonic background field. This creative work was afterwards developed by Unruh, Visser, Jacobson, Volovik, Garay, Schützhold, Fischer, Barceío, Balbinot et al [8-16]. Some of their works lead a more available way to experimental test on Hawking radiation by setting up a physical analogue gravity system, give probable models for gravity analogue in other objects, and open up many opportunities to study the gravity indirectly.

In this paper, we investigate the phonon radiation from a sonic black hole. The self-gravitation of the radiated phonons is taken into account. In the calculation, the method of Wentzel-Kramers-

---

[*]E-mail address: qdfang@163.com
[†]E-mail address: kaihu558895@sina.com

Brillouin ( WKB ) approximation is used. The feature of the potential barrier near the horizon and the process of the phonon tunnelling are described in detail. After calculating, we obtain the phonon emission rate and the radiation temperature which is in agreement with that derived by Unruh and Visser [7,10] by calculating the surface gravity.

## II. THE PHONON TUNNELLING FROM SONIC BLACK HOLE

Sound waves propagating in moving fluid will be dragged by the fluid flow, and if the speed of fluid exceeds the local sound speed, the sound waves can't go back upstream. This implies the sound waves are trapped in the supersonic flow region and a sonic black hole exists. Similar to the light waves in curved spacetime, the propagation of sound waves in barotropic, inviscid, and irrotational fluid flow can be determined by an effective metric (acoustic metric). For a spherically symmetric, stationary, convergent flow, the line element relating to sound propagating can be expressed as [10]

$$ds^2 = \frac{\rho_0}{c_s}\left[\left(c_s^2 - v^2\right)d\tau^2 - \frac{c_s^2}{c_s^2 - v^2}dr^2 - r^2\left(d\theta^2 + \sin^2\theta\, d\varphi^2\right)\right], \quad (1)$$

where $v$ is the velocity of the background fluid flow only with radial component, and $\rho_0$ denotes the density of the background flow, and $c_s$ is the local speed of sound. The boundary of the supersonic flow region, which locates at $r = R$ where $v$ is equal to $c_s$, is the horizon of the sonic black hole.

We can find that the line element in Eq. (1) is singular at $r = R$, and the existence of singularity at horizon brings troubles in trans-horizon physics. We can eliminate it by a coordinate transformation. Let $t = \tau + \int v/\left(c_s^2 - v^2\right)dr$, we have

$$ds^2 = \frac{\rho_0}{c_s}\left[\left(c_s^2 - v^2\right)dt^2 + 2v\,dr\,dt - dr^2 - r^2\left(d\theta^2 + \sin^2\theta\, d\varphi^2\right)\right]. \quad (2)$$

We can find that in the new coordinate system $(t, r, \theta, \varphi)$, the metric is regular at the horizon. And the metric keeps stationary, i.e. time-translation invariant, which manifests that the generator of $t$ is a killing vector. This coordinate covers the inside and outside of the sonic black hole.

### A. Radial phonon movement near the horizon

Expanding $v$ in the form of Taylor series with respect to $r$ at $r = R$ and notice that the flow moves inward, we get

$$v = -c_s + \alpha(r - R) + O\left((r - R)^2\right), \quad (3)$$

where $\alpha = \left.\frac{dv}{dr}\right|_{r=R}$. Neglecting the second and higher order terms in Eq. (3), the line element in Eq. (2) can be rewritten as

$$ds^2 = \frac{\rho_0}{c_s}\left[2\alpha c_s(r - R)dt^2 + 2\left(-c_s + \alpha(r - R)\right)dr\,dt - dr^2 - r^2\left(d\theta^2 + \sin^2\theta\, d\varphi^2\right)\right]. (4)$$

Now we study the speed of the outgoing phonon. The trajectory of a free phonon that moves along the radial direction is a sound-like geodesic line. So we have

$$ds^2 = \frac{\rho_0}{c_s}\left[2\alpha c_s (r-R)dt^2 + 2(-c_s + \alpha(r-R))dr\,dt - dr^2\right] = 0. \tag{5}$$

Solving the quadratic equation, we obtain the phonon speed

$$\dot{r} = \frac{dr}{dt} = -c_s + \alpha(r-R) \pm \sqrt{c_s^2 + \alpha^2(r-R)^2}. \tag{6}$$

The sign $\pm$ indicates the outgoing and ingoing phonon, respectively.

### B. WKB approximation

Similar to the case near the gravity horizon, because of the infinite blueshift near the sonic horizon observed from the infinity, we can get some reliable results in geometrical acoustics approximation ( WKB approximation ) without involving the second-quantized Bogoliubov method or the others. For a steady flow, the background keeps all the time, and the local fluid velocity $\upsilon$, the pressure and the density are constant at arbitrary space point. A radiating phonon in this spherically symmetric flow can be regarded as a $s$ wave. When a phonon is emitted from the sonic black hole, because of the total energy conservation, the energy inside the horizon will decrease. Then the position of the sonic horizon should shift from the initial place $r_i$ to the final place $r_f$. The initial position of the phonon corresponding to phonon pair creation is very close to the inner side of horizon $r_i$, and the final position of the phonon is nearly outside the final horizon $r_f$. Note how self-gravitation of the phonons is essential to the tunneling picture. Without self-gravitation, phonons created just inside the horizon would only have to tunnel just across an infinitesimal separation, so there would be no any barrier. But backreaction results in a shift of the horizon, it is the gap between $r_i$ and $r_f$ that forms the classical forbidden region, the potential barrier.

According to the WKB approximation, the $s$ wave of the radiated phonon can be expressed as $\Psi(r) = e^{iS(r)}$, where $S$ is the classical action. For an outgoing phonon across the horizon, the imaginary part of the action

$$\text{Im}\,S = \text{Im}\int_{r_i}^{r_f} p_r\,dr = \text{Im}\int_{r_i}^{r_f}\int_0^{p_r} dp_r'\,dr, \tag{7}$$

where $p_r$ is radial momentum of the outgoing phonons. Eliminating the momentum in favor of energy by using Hamilton's equation

$$\left.\frac{dH}{dp}\right|_r = \frac{\partial H}{\partial r} = \dot{r}, \tag{8}$$

where the Hamiltonian, $H$, is the generate of time $t$. When the horizon emits phonons with energy $\omega$, the remaining energy inside the sonic black hole becomes $H' = H - \omega$ because of the energy conservation. So, replacing $p_r$ with $\omega$ and note that $\omega$ ranges from 0 to $\omega$ in the tunneling process, we have

$$\mathrm{Im}\, S = \mathrm{Im} \int_{r_i}^{r_f} \int_0^\omega \frac{\mathrm{d}H'}{\dot{r}} \mathrm{d}r = -\mathrm{Im} \int_{r_i}^{r_f} \int_0^\omega \frac{\mathrm{d}\omega'}{\dot{r}} \mathrm{d}r. \tag{9}$$

Substitute Eq. (6) to Eq. (9) with signature $+$, we have

$$\mathrm{Im}\, S = -\mathrm{Im} \int_{r_i}^{r_f} \int_0^\omega \frac{\mathrm{d}\omega'}{-c_s + \alpha(r-R) + \sqrt{c_s^2 + \alpha^2(r-R)^2}} \mathrm{d}r, \tag{10}$$

then exchange the integral order and rearrange the expression as

$$\mathrm{Im}\, S = -\mathrm{Im} \int_0^\omega \mathrm{d}\omega' \int_{r_i}^{r_f} \left[ \frac{c_s + \sqrt{c_s^2 + \alpha^2(r-R)^2}}{2\alpha c_s (r-R)} - \frac{1}{2c_s} \right] \mathrm{d}r. \tag{11}$$

We can find that the integrand has a singularity at $r = R$ in the integral interval from $r_i$ to $r_f$. Utilizing the complex singular integral formula, we have

$$\mathrm{Im}\, S = \int_0^\omega \frac{\pi}{\alpha} \mathrm{d}\omega' = \frac{\pi\omega}{\alpha}. \tag{12}$$

Thus we get the phonon tunnelling probability, i.e., the rate of emission

$$\Gamma \sim \exp(-2\,\mathrm{Im}\, S) = e^{-2\pi\omega/\alpha}. \tag{13}$$

The equation above shows that the radiation spectrum is of black body spectrum with temperature

$$T_H = \frac{\hbar}{2\pi k_B} \alpha. \tag{14}$$

## CONCLUSION

For a sonic black hole, Hawking's formula was given by [7,10]

$$T_H = \left( \frac{\hbar}{2\pi K_B} \right) \frac{d(v-c)}{dx},$$

Where the surface gravity $\kappa$ in Hawking temperature $\kappa\hbar/2\pi K_B$ is replaced by $\frac{d(v-c)}{dx}$.

From Eq. (3), and notice the inverse of the sign between $c$ and $c_s$, we can get that $\frac{d(v-c)}{dx}$ just equals the constant $\alpha$. So we confirm the result we obtain in Eq. (14) agrees with that obtained by calculating the surface gravity [7, 10], the method we take is correct. More, the black body spectrum in Eq. (13) means the acoustical information lost completely. But if we allow for the higher orders in Eq. (3), the emission spectrum will deviate from exact black body spectrum. The correction of thermal spectrum implies the information at least partly conserved during the radiation of acoustic black hole.

# ACKNOWLEDGEMENT

This research is supported by the Natural Science Foundation of China (10847166).